\documentstyle[11pt]{article}
\begin{document}
\input epsf
\newcommand{\be}{\begin{equation}}
\newcommand{\ee}{\end{equation}}
\def\lqq{\lq\lq}
\def\rqq{\rq\rq}
\begin{titlepage}
%\today          \hfill
\begin{center}
\hfill    Preprint No BiBoS 720/1/96\\
\vskip .2in
{\large \bf Time of Events in Quantum Theory}\footnote
{This paper is dedicated to Klaus Hepp and to Walter Hunziker
on the occasion of their sixtieth anniversary}

\vskip .50in

Ph.~Blanchard${}^\flat$\footnote{
e-mail: blanchard@physik.uni-bielefeld.de} \ and\
A.~Jadczyk${}^\sharp$\footnote{
e-mail: ajad@ift.uni.wroc.pl}

{\em ${}^\flat$ Faculty of Physics and BiBoS,
University of Bielefeld\\
Universit\"atstr. 25,
D-33615 Bielefeld\\
${}^\sharp$ Institute of Theoretical Physics,
University of Wroc{\l}aw\\
Pl. Maxa Borna 9,
PL-50 204 Wroc{\l}aw}
\end{center}

\vskip 1.0in

\begin{abstract}
We enhance elementary quantum mechanics with three simple 
postulates that enable us to define time observable. We
discuss shortly justification of the new postulates and
illustrate the concept with the detailed analysis of
a delta function counter.  
\end{abstract}
\end{titlepage}
%\newpage
%\begin{quote}
\begin{flushright}
\begin{minipage}[t]{7cm}{\em Zeit ist nur dadurch, da\ss{} etwas geschieht und nur dort wo
etwas geschiecht.} \\(E.Bloch)
\end{minipage}
%{\em Zeit ist nur dadurch, da\ss{} etwas geschieht\\
% und nur dort wo 
%etwas geschiecht.
%}(E.Bloch)\\
\end{flushright}
%\end{quote}
\section{Introduction}
Time plays a peculiar role in Quantum Mechanics. It differs from other
physical quantities like position or momentum. When discussing
position a dialogue may look like this:
\footnote{We use the method chosen by Galileo in his great book "Dialogues
Concerning Two New Sciences"\cite{galileo}. Galileo is often refered to as the
founder of modern physics. The most far--reaching of his achievements was his
counsel's speech for mathematical rationalism against Aristotle's
logico--verbal approach, and his insistence for combining mathematical
analysis with experimentation.\\ In the dialog SG=Sagredo,
SV=Salviati,
SP=Simplicio}\\

\indent SP: What is the position?\\
\indent SG: Position of what?\\
\indent SP: Of the particle.\\
\indent SG: When?\\
\indent SP: At t=t1.\\
\indent SG: The answer depends on how  you are going to measure this position. 
Are
you sure you have detectors put everywhere that interact with the
particle only during the time interval (t-dt,t+dt) and not before?\\

\noindent
When talking about time we will have something like this:\\

SP: What is time?\\
\indent SG: Time of what?\\
\indent SP: Time of a particle.\\
\indent SG: Time of your particle doing what?\\
\indent SP: Time of my particle leaving the box where it was trapped. Or
time at which my particle enters the box.\\
\indent SG: Well, it depends on the box and it depends on the method
you want to apply to ascertain that the event has happened.\\
\indent SP: Why can't we simply put clocks everywhere, as it is common in
discussions of special relativity? And let these clocks note
the time at which the particle passes them?\\
\indent SG: Putting clocks disturbs the system. The more clocks you put -
the more you disturb. If you put them everywhere - you force
wave packet reductions a`la GRW. If you increase their time
resolution more and more - you increase the frequency of
reductions. When the clocks have infinite resolutions - then
the particle stops moving - this is the Quantum Zeno effect \cite{blaja93b}.\\
\indent SP: I do not believe these wave packet reductions. Zeh published a
convincing paper whose title tells its content: "There are no
quantum jumps nor there are particles" \cite{zeh93}, and Ballentine
\cite{ball90,ball91}, proved
that the projection postulate is wrong.\\
\indent SG I remember these papers. They had provocative titles...\\
\indent SV.  First of all Ballentine did not claim that the projection
postulate is wrong. He said that if incorrectly applied - then
it leads to incorrect results. And indeed he showed how incorrect
application of the projection postulate to the particle tracks in a cloud
chamber leads to inconsistency.  What he said is this:
"According to the projection postulate, a position measurement should
"collapse" the state to a position eigenstate, which is spherically
symmetric and would spread in all directions, and so there would be no
tendency to subsequently ionize only atoms that lie in the direction of
the incident momentum. An approximate position measurement
would similarly yield a spherically symmetric wave packet,
so the explanation fails." This is exactly what he said. And this
is correct. This shows how careful one has to be with the projection
postulate.
If the projection postulate is understood as operating with an
operator on a state vector: $\psi\longrightarrow R\psi/\Vert R\psi\Vert$,
then the argument does not apply. Thus
a correct application would be to  {\em multiply} the moving Gaussian
of the particle, something like:
$$\psi(x,t)=\exp(i{\bf p}({\bf x}-{\bf a}(t))\exp(-({\bf x}-
{\bf a}(t))^2/\sigma(t))$$
which is spherically symmetric, but only up to the phase,
by a static Gaussian modelling  a detector localized around $a$:
$$f(x)=\exp(-\alpha ({\bf x}-{\bf a})^2)$$
The result is again
a moving Gaussian. And in fact, such a projection postulate is not
a postulate at all. It can be derived from the simplest possible
Liouville equation.\\
\indent SP: Has this "correct", as you claim, cloud chamber model been
published? Have its predictions been experimentally verified?\\
\indent  SV: A general theory of coupling between quantum system
and a classical one is now rather well understood \cite{blaja95a}.
The cloud chamber model has been published quite recently, you
can take a look at \cite{jad94b,jad94c}.
Belavkin and Melsheimer \cite{belmel}
tried to derive somewhat similar result from a pure unitary
evolution, but I am not able to say what assumptions they used,
what approximations they made and what exactly are their results.\\
\indent SP: Hasn't the problem been solved long ago in the classical
paper by Mott \cite{mott}?\\
\indent SV
Mott did not even attempt to derive the timing of the tracks.
In the cloud chamber model
of Refs. \cite{jad94b,jad94c},
that I understand rather well, because I participated in its construction,
it is interesting that the detectors -- even if they do not "click" --
influence the continuous evolution of the wave packet between
reductions. They leave a kind of a "shadow".
This is another case of a "interaction-free"
experiment discussed by Dicke \cite{dicke81,dicke88}, and then by Elitzur and
Vaidman \cite{eliva} in their "bomb--test" allegory,
and also by Kwiat, Weinfurter, Herzog and
Zeilinger \cite{kwiat}. The shadowing effect predicted by EEQT
\footnote{Salviatti refers here to "Event Enhanced Quantum
Theory" of reference \cite{blaja95a} - paper apparently well know to
the participants of the dialog.}  may be tested
experimentally. I believe it will find many applications in
the future, and I hope these will be not only the military ones!
Yet
we must now not digress upon this particular topic since you are
waiting to hear what I think about the problem of time in quantum theory.
We already know that "time" must be "time of something". Time of
something that happens. Time of some event.\footnote
{Heisenberg proposed the word "event" to replace the word
"measurement", the latter word carrying a suggestion of 
human involvement.} But in quantum
theory events are not simply space-time events as it is in
relativity. Quantum theory is specific in the sense that there
are no events unless there is something external to the quantum
system that "causes" these events. And this something external must
not be just another quantum system. If it is just another quantum system -
then nothing happens,
only the state vector continuously evolves in parameter time.\\
\indent SP But is it not so that there are no sharp events?
Nothing is sharp, nothing really sudden. All is continuous. All
is approximate.\\
\indent SG How nothing is sharp, do we not register "clicks"
when detecting particles?\\
\indent SP I do not know what clicks are you talking about ...\\
\indent SG How you don't know? Ask the experimentalist.\\
\indent SP I {\em am} an experimentalist!\\
\indent SV The problem you are discussing is not an easy
one to answer. I pondered on it many times, but did not
arrive at a clear conclusion. Nevertheless something can
be said with certainty. First of all you both agree that
in physics we always have to deal with idealizations.
For instance one
can argue that there are no real numbers, that the only,
so to say, experimental numbers are the natural numbers.
Or at most rational numbers. But real numbers proved to be useful
and today we
are open to both possibilities: of a completely discrete
world, and of a continuous one. Perhaps there is also a third
possibility: of a fuzzy world. Similarly there are
different options for modeling the events. One can
rightly argue that they are never sharp. But do they
happen or not? Do we need counting them? Do we
need a theory that describes these counts? We do.
So, what to do? We have no other choice but to try
different mathematical models and to see which of
them better fit the experiment, better fit the rest
of our knowledge, better explain what makes Nature tick.
In the cloud chamber model that
we were talking about just a while ago the events
are unsharp in space but they are sharp in time. And the
model works quite well. However, if you try to work out
a relativistic cloud chamber model, then you see
that the events must be also smeared out in the
coordinate time.\footnote
{Cf. an illuminating discussion of this point
in \cite{unruh90,birula94}.} Nevertheless they can still 
be sharp in
a different "time", called "proper time" after Fock
and Schwinger. If time allows I will tell you
more about this relativistic theory, but now let
us agree that in a nonrelativistic theory sharp
localization of events in time does not contradict
any known principles. We will remember at the same
time that we are dealing here with yet another
idealization that is good as long as it works
for us. And we must not hesitate to abandon it
the moment it starts to lead us astray. The principal
idea of EEQT is the same as that expressed in a recent
paper by Haag \cite{haag96}. Let me quote only this:
"... we come almost unavoidably to an evolutionary
picture of physics. There is an evolving pattern of events 
with causal links connecting them. At any stage the `past'
consists of the part which has been realized, the `future'
is open and allows possibilities of new events ..."\\
\indent SG Let me interrupt you. Perhaps we
should remember what Bohr was telling us. Bohr insisted
that the apparatus has to be described in terms of
classical physics; this point of view is a
common--place for experimental physicists. Indeed
any experimental article observes this rule. This
principle of Bohr is not in any way a contradiction but
simply the recognition of the fact that any physical
theory is always the expression of an approximation
and an idealization. Physics is always a little bit
false. Epistemology must also play role in the labs.
Physics is a system of analogies and metaphors. But
these metaphors are helping us to understand how
Nature does what it does.\\
\indent SP I agree with this. So what is your
proposal? How to describe time of events in
a nonrelativistic quantum theory? Does one first
have to learn EEQT - your "Event Enhanced
Quantum Theory" that you are so proud of?
I know many theoretical physicists dislike your explicit
introduction of a classical system. They prefer
to keep everything classical in the background.
Never put it into the equations.\\
\noindent SV Here we have a particularly lucky
situation. For this particular purpose of discussing
time of events it is not necessary to learn EEQT.
It is possible to describe time observation with
simple rules. This is normal in standard quantum
mechanics. You are told the rules, and you are told
that they work. So you believe them and you are happy
that you were told them. In EEQT
Schr\"odinger's evolution and reduction of the wave function
appear as special cases of a single law of motion which
governs all systems equally. EEQT is one
of the few approaches that allow you to {\sl derive}
quantum mechanical postulates and to see that these postulates
reflect only part of the truth. Here, when discussing
time of events we do not need the full predictive
power of EEQT. This is so because after an event has been
registered the experiment is over. We are not
interested here in what happens to our system after
that. Therefore we need not to speak about jumps and
wave packet reductions. It is only  if you want to
{\bf derive} the postulates for time measurements, only
then you will have to look at EEQT. But instead of deriving the rules,
it is even better to see if they give experimentally
valid predictions. We know too many cases where
good formulas were produced by doubtful methods and
bad formulas with seemingly good ones. Using the right tool 
makes the job easier.\\
\indent SP I become impatient to see your postulates,
and to see if I can accept them as reasonable even
before any experimental testing. Only if I see that they
are reasonable, only then I will have any motivation to
see whether they really be derived from EEQT, or perhaps
in some other way.

\section{Time of Events}
We start our discussion on quite a general and somewhat abstract
level. Only later on, in examples, we will specialize
both: our system and the monitoring device.
We consider quantum system described using a Hilbert
space ${\cal H}$.\footnote{More 
generally we would need two Hilbert spaces:
${\cal H}_{no}$ and ${\cal H}_{yes}$ that can be different, 
but for the present discussion we need not be pedantic, so we 
will assume them to be identified.} 
 To answer the question "time of what?",
we must select a {\sl property} of the system that we are going
to monitor. It must give only "yes-no", or {\sl one-zero} answers.
We denote this binary variable with the letter $\alpha$. In our
case, starting at $t=0$,
 when the monitoring begins, we will get continuously $\alpha=0$ reading
on the scale,
until at a certain time, say $t=t_1$, the reading will change into "yes".
Our aim is to get the statistics of these "first hitting times",
and to find out its dependence  of the initial state of
the system and on its dynamics.\\
Speaking of the "time of events" one can also think that
"events" are transitions which occur; sometimes the system
is changing its state randomly - and these changes are
registered. There are two kinds of probabilities in
Quantum Mechanics the transition probabilities and
other probabilities - those that tell us {\bf when} the
transitions occur. It is this second kind of probabilities
that we will discuss now.
\subsection{First Postulate - the Coupling}

Our first postulate reads: the coupling to a "yes-no"
monitoring device is described by an operator
$\Lambda\geq 0$ in the Hilbert space ${\cal H}$.
In general  $\Lambda$ may explicitly
depend on time but here, for simplicity, we will assume that
this is not the case. That means: to any real experimental
device there corresponds a $\Lambda$. In practice it may be
difficult to produce the $\Lambda$ that describes exactly
a given device. As it is difficult to find the Hamiltonian
that takes into account {\sl exactly}\ all the forces that
act in a system. Nevertheless we believe that an exact
Hamiltonian exists, even if it is hard to find or impractical
to apply. Similarly our postulate states that an exact $\Lambda$
exists, although it may be hard to find or impractical
to apply. Then we use an approximate one, a model one.

It should be noticed that we do not assume that $\Lambda$
is an orthogonal projection. This reflects the fact that our
device  - although giving definite "yes-no" answers,
gives them acting upon possibly {\sl fuzzy}\ criteria. In the
limit when the criteria become sharp one should think
of $\Lambda$ as $\Lambda\longrightarrow \lambda E$, where
$\lambda$ is a coupling constant of physical dimension
$t^{-1}$ and $E$ a projection operator. In the general case
it is usually convenient to write $\Lambda=\lambda \Lambda_0 ,$
where $\Lambda_0$ is dimensionless.

It is also important to notice that the property that is being
monitored by the device
need not be an {\sl elementary} one. Using the concepts
of quantum logic (cf. \cite{jauch,piron}) the property need not 
be {\sl atomic} -
it can be a composite property. In such a case,
when thinking about physical implementation of the procedure
determining whether the property holds or no, there are two
extreme cases. Roughly speaking the situation here is similar to
that occurring in discussion of superpositions of state preparation
procedures. Some procedures lead to a coherent superpositions,
some other lead to mixtures. Similarly with composite detectors:
 one possibility is that we have
a distributed array of detectors that can act independently of each
other, and our event consists
on activating {\em one of them}. Another possibility is
that we have a {\sl coherent}\ distributed detector like a
solid state lattice that acts as {\em one} detector.
In the first case (called {\sl incoherent}) $\Lambda$ will
be of the form
$$\Lambda=\sum_\alpha g_\alpha^{\star} g_\alpha ,$$
while in the second {\sl coherent} case:
$$\Lambda=(\sum_\alpha g_\alpha)^{\star}(\sum_\alpha g_\alpha),$$
where $g_\alpha$ are operators associated to individual
constituents of the detector's array. More can be said
about this important topic, but we will not need to
analyze it in more details for the present purpose.

\subsection{Second Postulate - the Probability}
We assume that, apart of the monitoring device, our system
evolves under time evolution described by the Schr\"odinger
equation with a self--adjoint Hamiltonian $H_0=H_0^\star$. We denote by
$K_0(t)=\exp (-iH_0 t)$ the corresponding unitary propagator.
Again, for simplicity, we will assume that $H_0$ does not
depend explicitly on time.

Our second postulate reads: assuming that the monitoring
started at time $t=0$, when the system was described
by a Hilbert space vector $\psi_0$, $\Vert\psi_0\Vert=1$,
 and when the monitoring
device was recording the logical value "no", the
probability $P(t)$ that the change $no\rightarrow yes$ will
happen before time $t$ is given by the formula:
\be
P(t)=1-\Vert \psi_t\Vert^2,
\label{eq:pt}
\ee
where
\be
\psi_t=K(t)\psi_0
\label{eq:psit}
\ee
and
\be
K(t)=\exp(-iH_0 t - {{\Lambda t}\over2}).
\label{eq:kt}
\ee
{\bf Remark}: The factor ${1\over2}$ in the formula above is put
here for consistency with the notation used in our previous
papers.

It follows from the formula (\ref{eq:pt}) that the probability
that the counter will be triggered out in the time interval (t,t+dt),
provided it was not triggered yet, is $p(t)dt$, where $p(t)$ is
given by
\be
p(t)={d\over{dt}}P(t)=<K(t)\psi_0,\Lambda K(t)\psi_0>.
\label{eq:pst}
\ee
We remark that $\int_0^\infty p(t)dt = P(\infty)$ is the probability that
the detector will notice the particle at all. In general this number
representing the total efficiency of the detector (for a given initial
state) will be smaller than $1.$

\subsection{Third Postulate - the Shadowing Effect}
As noticed above in general we expect $P(\infty)<1$. That means
that if the experiment is repeated many times, then there will be
particles that were not registered while close to the counter;
they moved away, and they will never be registered in the future.
The natural question then arises: is the very presence of the
counter reflected in the dynamics of the particles that pass
the detector without being observed? Or we can put it as a
"quantum espionage" question: {\bf can a particle detect a detector
without being detected?} And if so - which are the precise
equations that tell us {\bf how}?.

To answer this question it is not enough to use the two postulates
above. One needs to make use of the Event Engine of EEQT once more.

\begin{figure}[t]
\epsfysize=18cm
\epsffile{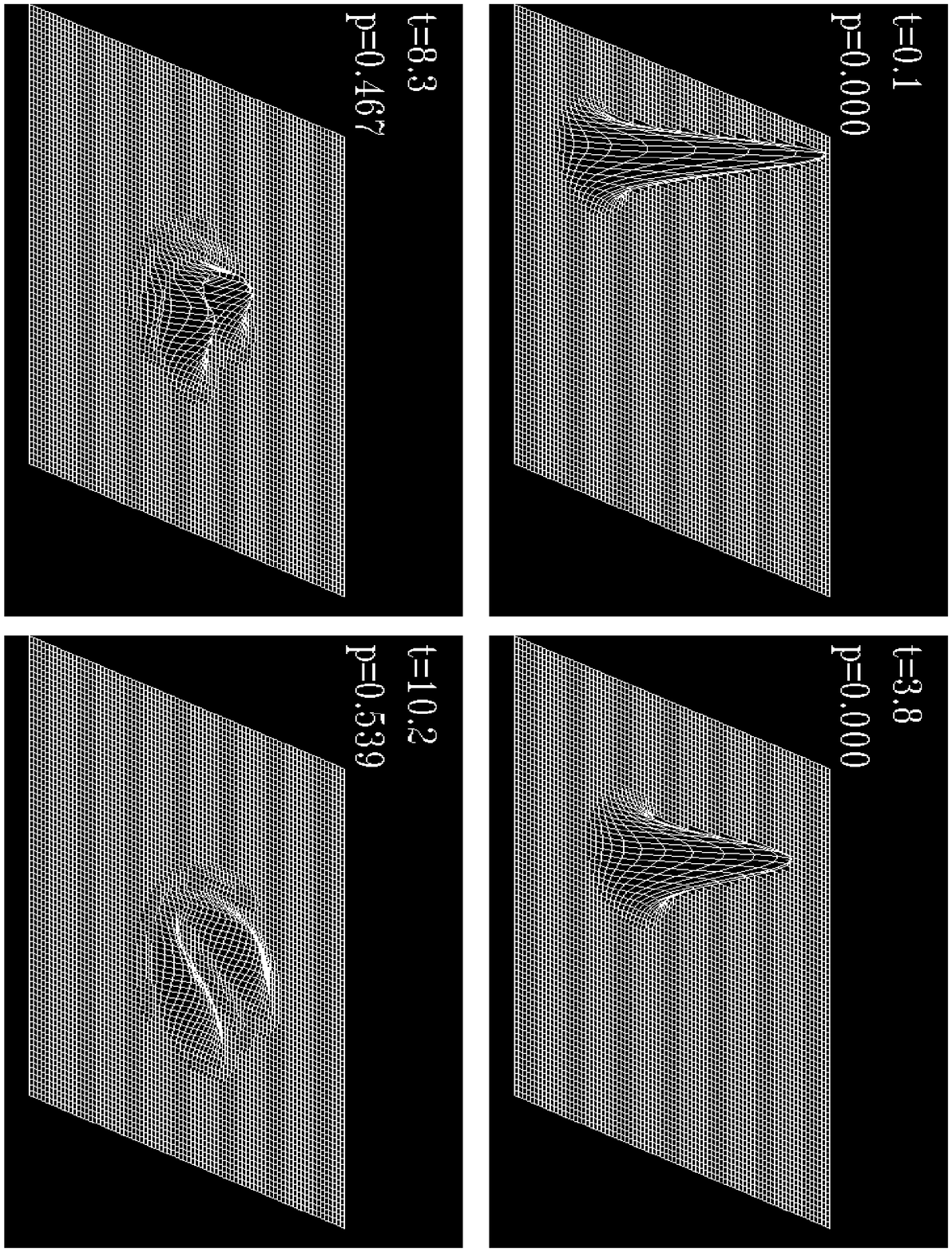}
\caption{Four shots from the time evolution of a
gaussian wavepacket monitored by a gaussian detector placed
at the center of the plane. The efficiency of the detector is
in this case ca. $P(\infty) \simeq 0.55 .$ }
\label{fig:fig1}
\end{figure}

Our third postulate reads:
prior to any event, and independently of whether any event will
happen or not, the state of the system is described by the vector
$\psi_t$  undergoing the non unitary
evolution given by Eq. (\ref{eq:psit}). It is not too
difficult to think of an experiment that will test this
prediction.\\
Fig. \ref{fig:fig1} shows four shots from time evolution of a
gaussian wavepacket monitored by a gaussian detector placed
at the center of the plane. The efficiency of the detector is
in this case ca. $P(\infty )\simeq 0.55 .$ There is almost no
reflection. The shadow of the detector that is seen on the 
fourth shot can be easily interpreted in terms of ensemble
interpretation: once we count only those particles that were
{\em not}\ registered by the detector, then it is clear that
there is nothing or almost nothing behind the detector. 
However a careful observer will notice that there is
a local maximum exactly behind the counter. This is 
a quantum effect, that of "interference of alternatives". 
It has consequences for the rate of future events for 
an individual particle.
\vspace{10pt}

\subsection{Justification of the postulates}
The above postulates are more or less "natural". They are in agreement
with the existing ideas of non-unitary\footnote{Known in the literature
also under the name of "non-hermitian"} evolution. So, for
instance, in \cite{mignani90} the authors considered the ionization
model. They wrote: \lq According to the usual procedure the
ionization probability $P(t)$ should be given by
$P(t)=1-\vert \Psi \vert^2$\rq .\\
Even if our postulates are
natural, it is worthwhile to notice that EEQT allows
us to interpret them, to understand them and to derive them, in terms of
{\em classical} Markov processes.
First of all let us see that the above formula for $P(t)$ can be understood
in terms of an inhomogeneous Poisson decision process as follows.\footnote
{A mathematical theory of a counter that leads to an inhomogeneous
Poisson process,  starting from formal postulates that 
are different than ours was given almost fifty 
 years ago by Res Jost \cite{jost46} .} Assume
the evolution starts with  some quantum state $\psi_0$,  of norm one, as
above. Define the positive function $\lambda(t)$ as
\be
\lambda(t)=({\hat \psi_t},\Lambda {\hat \psi_t}),
\label{eq:lambda}
\ee
where
\be
{\hat \psi_t}={{\psi_t}\over {\Vert \psi_t\Vert}},
\label{eq:hatpsit}.
\ee
Then $P(t)$ above happens to be nothing but the first-jump probability
of the inhomogeneous Poisson process with intensity $\lambda(t)$.
It is instructive to see that this is indeed the case. To this
end let us divide the interval $(0,t)$ into $n$ subintervals
of length $\Delta t = t/n$. Denote $t_k=(k-1) \Delta t ,$ $k=1,\ldots,n .$
The inhomogeneous Poisson process of intensity $\lambda (t)$ consists then of
taking independent decisions \lq jump--or--not--jump\rq\
in each time subinterval.
The probability  for jumping in the $k$-th subinterval is assumed to be
$p_k=\lambda(t_{k-1})\Delta t$ (that is why $\lambda$ is called the
{\sl intensity}\
of the process).
Thus the probability $P_{not}(t)$ of not jumping up to time $t$ is
\be
P_{not}(t)=\lim_{n\rightarrow\infty} \prod_{k=1}^n (1-p_k)=\exp ( -\int_0^t
\lambda(s) ds ).
\label{eq:p1t}
\ee
Let us show that $1-P_{not}(t)$ can be identified with $P(t)$ given by Eq. 
(\ref{eq:p1t}).
To this end notice that
\begin{eqnarray}
\frac{d}{dt} (1-P(t))&=&- <\psi_t ,\Lambda \psi_t >=
-\lambda(t) \Vert \psi_t\Vert^2\nonumber\\
&=&-\lambda(t) (1-P(t)).\nonumber
\end{eqnarray}
Thus $1-P(t)$, given by Eq. (\ref{eq:pt}) satisfies the same differential 
equation as
$P_{not}(t)$ given by Eq. (\ref{eq:p1t}). Because $1-P(0)=P_{not}(0)=1$, it
follows that $1-P(t)=P_{not}(t)$, and so $P(t)=1-P_{not}(t)$ indeed is the first 
jump
probability of the inhomogeneous Poisson process with intensity $\lambda (t)$.

This observation is useful but rather trivial. It can not yet stand
for a justification of the formula (\ref{eq:pt}) - this for the simple
reason that the jump process above, based upon a continuous observation
of the variable $\alpha$ and registering the time instant of its jump,  is not
a Markovian process. It would become
Markovian if we know $\lambda(t)$, but to know $\lambda(t)$ we must
know $\psi_t$. This leads us to consider pairs $x_t=(\psi_t,\alpha_t)$, 
where
$\psi_t$ is the Hilbert space vector describing quantum state, and $\alpha_t$
is the yes-no state of the counter. Then $\psi_t$ evolves deterministically
according to the formula (\ref{eq:psit}), the intensity function $\lambda(t)$
is computed on the spot, and the Poisson decision process described above
is responsible for the jump of value of $\alpha$ - in our case
it corresponds to a "click" of the counter. The time of the click is a
random variable $T_1$, well defined and computable by the above prescription.\\
This prescription sheds some  light onto the meaning of the quantum state vector $\psi$.
We see that $\psi$ codes in itself information that is being used by
a decision mechanism working in an entirely classical way - the
only randomness we need is that of a biased (by $\lambda(t)$)
classical roulette. Until we ask {\em why} the bias is determined
by this particular functional of the quantum state, until then we
do not have to invoke more esoteric concepts of quantum probability
-- whatever they mean.
But, in fact, it is possible to understand somewhat more, still in pure
classical terms. We will not need this extra knowledge in the rest
of this paper, but we think it is worthwhile to sketch here at least
the idea.\\
In the reasoning above we were interested only in what governs the
time of the first jump, when the counter clicks. But in reality
nothing ends with this click. A photon, for instance, when detected,
is transformed into another form of energy. So, if we want to
continue our process in time, after $T_1$, we must feed it
with an extra information: how is the quantum state transformed
as the result of the jump. So, in general, we have a classical
variable $\alpha$ that can take finitely many, denumerably many,
a continuum, or more, possible values, and to each ordered pair
$(\alpha\rightarrow\beta )$ there corresponds an operator
$g_{\alpha\beta}$. The transition $(\alpha\rightarrow\beta )$
is called an {\em event}, and to each event there corresponds
a transformation of the quantum state
$\psi\rightarrow \frac{g_{\beta\alpha}\psi}{\Vert g_{\beta\alpha}
\psi\Vert}$.
In the case of a counter there is only one $\beta$. In general, when
there are several $\beta$-s,  we need  to tell not only when to jump,
but also where to jump. One obtains in this way a piecewise deterministic
Markov process on pure states of the total system: (quantum object,
classical monitoring device). It can be then shown \cite{blaja95a,jakol95}
that this process, including
the jump rate formula (\ref{eq:lambda}) follows uniquely from the simplest
possible Liouville equation that couples the two systems.

\section{The Time of Arrival}
As the most natural application of the above concept of "time of event"
we consider the notion of "time of arrival" of a particle to a certain
state. There are several
methods available for computing the "time of arrival" distribution
given our postulates. We shall take the approach that seems to us
to be the simplest one. One by one we shall specialize our
assumptions about $\Lambda$.

\subsection{One elementary detector}
Let $K(t)$ be given by Eq. (\ref{eq:kt}), and let\footnote{From this
time on the subscript $_0$ may refer to either the initial state, as
in $\psi_0$ or to free evolution, as in $H_0$, or to initial state 
evolving under free evolution. In case of confusion the actual meaning 
should be derived from the context.}
\be
K_0(t)=\exp (-iH_0t).
\ee
Then $K(t)$ satisfies the Schr\"odinger equation
\be
{\dot K}=-iH_0 K(t)-{\Lambda\over 2} K(t) .
\ee

This differential equation, together with initial data $K(0)=I$, is easily seen to be
equivalent to the following integral equation:
\be
K(t)=K_0(t)-{1\over 2}\int_0^t K_0(t-s)\Lambda K(s)ds .
\ee
By taking the Laplace transform and by the convolution theorem we
get the Laplace transformed equation:
\be
{\tilde K}={\tilde K}_0-{1\over 2}{\tilde K}_0\Lambda{\tilde K}.
\label{eq:klap}
\ee
Let us consider the case of a maximally sharp measurement. In this case
we would take $\Lambda=|a><a|$, where $|a>$ is some Hilbert space vector.
It is not assumed
to be normalized; in fact its norm stands for the strength of the coupling
(notice that $<a|a>$ must have physical dimension $t^{-1}$).
Taking look at the formula (\ref{eq:pst})
we see that now $p(t)=\vert<a|\psi_t>\vert^2$ and so we need to know
$<a|K(t)\psi_0>$
rather than the full propagator $K(t)$. Multiplying Eq. (\ref{eq:klap})
from the left by $<a|$ and from the right by $|\psi_0>$ we obtain:
\be
<a|{\tilde\psi}>={2<a|{\tilde K_0}|\psi_0>
\over{2+{<a|{\tilde K}_0|a>}}}
\label{eq:atp}
\ee
where ${\tilde \psi}$ is the Laplace transform of $t\rightarrow \psi(t)$ :
\be
{\tilde \psi}(z)=\int_0^\infty e^{-tz} \psi(t) dt = {\tilde K}(z)\psi_0, 
\qquad \Re(z)\geq 0 .
\label{eq:psiz}
\ee
\subsection{Composite detector}
We consider now the simplest case of a composite detector. It will be
an incoherent composition of two simple ones. Thus we will take:
\be
\Lambda=|a_1><a_1|+|a_2><a_2| .
\ee
{\bf Remark} Notice that if $<a_1|a_2>=0$,  then coherent and incoherent
compositions are indistinguishable, as in this case, with $g_i=|a_i><a_i| ,$
 we have that $\sum_i {g_i}^\star g_i = (\sum_i {g_i})^\star
(\sum_i {g_i}).$\\
For $p(t)$ we have now the formula:
\be
p(t)=\sum_i \vert <a_i|\psi_t>\vert^2,
\ee
and to compute the complex amplitudes $<a_i|\psi_t>$ we
will use the Laplace transform method as in the case of one detector.
To this end one applies $<a_i|$ from the left and $|\psi_0>$
from the right to Eq. (\ref{eq:klap}) and solves the resulting system
of two linear equations to obtain:
\begin{eqnarray}
\left.
\begin{array}{ll}
<a_1|{\tilde\psi}>&={2\over{\Delta}}\ (
(2+(22))<a_1|{\tilde\psi}_0>-(12)<a_2|{\tilde\psi}_0>\ )
\\
&\\
<a_2|{\tilde\psi}>&={2\over{\Delta}}\ (
(2+(11))<a_2|{\tilde\psi}_0>-(21)<a_1|{\tilde\psi}_0>\ )
\end{array} \right\}
\end{eqnarray}
where we used the notation
\be
(ij)=<a_i|{\tilde K_0}|a_j>,
\ee
\be
|{\tilde\psi}_0>={\tilde K_0}|\psi_0> ,
\ee
and where $\Delta$ stands for
\be
\Delta= 4+2((11)+(22))+((11)(22) - (12)(21) ) .
\ee
The probability density $p(t)$ is then given by
\be
p(t)=\sum_i \vert \phi_i (t)\vert^2,
\label{eq:pti}
\ee
where $\phi_i$ is the inverse Fourier transform
\be
\phi_i (t)={1\over{2\pi}}\int_{-\infty}^\infty e^{ity} {\tilde \phi}_i (iy) dy 
\label{eq:fourier}
\ee
of
\be
{\tilde \phi}_i(iy)=\lim_{x=0+} <a_i|{\tilde \psi}(x+iy)>\ .
\ee 

By the Parseval formula we have
that $P(\infty)$ is given by:
\be
P(\infty)={1\over{2\pi}}\sum_i \int_{-\infty}^{\infty} 
\vert {\tilde \phi}_i (iy)\vert^2 dy \ .
\ee
\subsection{Example: Dirac's $\delta$ counter for ultra-relativistic particle}
Let us now specialize the model by assuming that we consider a particle
in ${\bf R}^1$ and that the Hilbert space vector $|a>$ approaches the
improper position eigenvector ${\sqrt \kappa}\delta(x-a)$ localized at the point $a$.
This corresponds to a point--like detector of strength $\kappa$ placed at $a$.\footnote
{The case of Hermitian singular $delta$--function perturbation was discussed
by many authors - see \cite{bauch85,gav85,blinder88,lav88,albev88,manoukian89,
grosche90}
and references therein}
We see from the
equation (\ref{eq:pst}) that $p(t)$ is in this case given by:
\be
p(t)=|\phi (t)|^2 ,
\ee
where the complex amplitude $\phi(t)$ of the particle arriving at $a$ is:
\be
\phi (t)=<a|\psi (t)>,
\ee
or, from Eq. (\ref{eq:atp})
\be
{\tilde \phi}=
{{2\sqrt{\kappa}}\over{2+{\kappa}{\tilde K}_0(a,a)}}{\tilde \psi}_0(a)
\label{eq:ak}
\ee
where
${\tilde \psi}_0$ stands for the Laplace transform of $K_0(t)\psi_0$.\\
Let us now consider the simplest explicitly solvable example - that of an
ultra--relativistic particle on a line. For $H_0$ we take $H_0=-ic{d\over{dx}} ,$
then the propagator $K_0$ is given by $K_0(x,x';t)=\delta(x'-x+ct) ,$ and its
Laplace transform reads ${\tilde K}_0(x,x';z)={1\over c}e^{(x-x')z/c}$. In
particular ${\tilde K}_0(a,a;z)={1\over c}$ and from Eq. (\ref{eq:ak}) we
see that the amplitude for arriving at the point $a$ is given by the "almost
evident" formula:
\be
\phi(t)=const(\kappa)\times\psi(a-ct) ,
\ee
where
$const(\kappa)={\sqrt \kappa}/(1+{\kappa\over{2c}}) .$
It follows that probability that the particle will be registered is equal to
\be
P(\infty)={{\kappa/c}\over{(1+{\kappa\over{2c}})^2}}\int_{-\infty}^{a}dx\vert
\psi_0(x)\vert^2
\ee
which has a maximum $P(\infty)=1/2$ for $\kappa=2c$ if the support of $\psi_0$
is left to the counter position $a .$ We notice that in this
example the
shape of the arrival time probability distribution $p(t)$ does not depend
on the value of the coupling constant - only the effectiveness of the
detector depends on it. For a counter corresponding to a superposition
$\sum_i \sqrt{\kappa_i}\delta(x-a_i)$ we obtain for $P(\infty)$ exactly the
same expression as for one counter but with $\kappa$ replaced with 
$\sum_i\kappa_i .$

\subsection{Example: Dirac's $\delta$ counter for Schr\"odinger's particle}
We consider now another example corresponding to a free Schr\"odinger's particle on a line. 
We will study response of a Dirac's delta counter $|a>=\sqrt{\kappa}\delta (x-a)$,
 placed at $x=a$, to a Gaussian wave packet whose initial shape
 at $t=0$ is given by:
\be
\psi_0(x)=
{
1
\over
{
(2\pi )^{1/4}\eta^{1/2}
}
}\exp 
\left(
{{ -(x-x_0)^2}\over{4\eta^2}}
+2ik(x-x_0)
\right) .
\ee
In the following it will be convenient to use dimensionless variables
for measuring space, time and the strength of the coupling:
\be
\xi={x\over 2\eta},\qquad \tau={{\hbar t}\over{2m\eta^2}},\qquad
\alpha={{m\eta\kappa}\over\hbar}
\ee
We denote 
\be
\xi_0=x_0/2\eta ,\; \xi_a=a/2\eta ,\; v=2\eta k  
\ee
In these new variables we have:
\begin{eqnarray}
\psi_0 (\xi)=&
\left(
{2\over\pi}\right)^{1/4}
e^{-(\xi-\xi_0)^2+2iv(\xi-\xi_0)}\\
K_0(\xi',\xi;\tau)=&\left(
{1\over{\pi i\tau}}
\right)^{1/2}\exp \left( {{i (\xi'-\xi)^2}
\over\tau}\right)\\
{\tilde K}_0 (\xi',\xi;z)=&(iz)^{-{1\over2}} 
\exp \left( -2\sqrt{-iz}\, \vert\xi'-\xi\vert\right)
\end{eqnarray}
We can compute now explicitly ${\tilde \psi}(z)$ of
Eq. (\ref{eq:psiz}):
\be
{\tilde\psi}_0(a;z)={1\over2}(2\pi)^{1/4}(iz)^{-1/2}e^{-d^2-2ivd}
\left[ \mbox{w}(u_+) +\mbox{w}(u_-)\right]
\ee
where
\be
u_{\pm}=i\sqrt{-iz}\, \pm(v-id) , \qquad d=\xi_0-\xi_a ,
\ee
and the amplitude ${\tilde\phi}$ of Eq. (\ref{eq:ak}), when rendered
dimensionless,\footnote{We should have $\vert\phi(t)\vert^2 dt=
\vert\phi(\tau)\vert^2d\tau $}  reads
\be
{\tilde\phi}(z)={1\over2}(2\pi)^{1/4}\alpha^{1/2}e^{-d^2-2ivd}\, 
\frac{\mbox{w}(u_+) +\mbox{w}(u_-)}
{2\sqrt{iz}\, +\alpha}
\ee
with the function $w(z)$ defined by
\be
\mbox{w}(u)=e^{-u^2}\, \mbox{erfc}(-iu)
\ee
(see Ref. \cite{abram}, Ch. 7.1.1 -- 7.1.2). We have also used
the formula
\be
\int_0^\infty e^{-(ax^2+2bx+c)}dx=
{1\over 2}\sqrt{{\pi\over a}}\exp\left(\frac{b^2-ac}{a}\right)\mbox{erfc}
\left(\frac{b}{\sqrt{a}}\right)
\ee
valid for $\Re(a) > 0$ (see \cite{abram}, Ch. 7.4.2).\\
To compute $p(t)$ from Eqs. (\ref{eq:pti},\ref{eq:fourier}) the correct boundary values
of the complex square root (with the cut on the negative real half-axis)
must be taken. Thus for $z=x+iy,\, x=0+$ we should take
\begin{eqnarray}
\sqrt{iz}=
\left\{
\begin{array}{ll}
i\sqrt{y}\qquad &y\geq0\\
\sqrt{-y}\qquad &y\leq0
\end{array}
\right.
\end{eqnarray}
\begin{eqnarray}
\sqrt{-iz}=\left\{
\begin{array}{ll}

\sqrt{y}\qquad &y\geq0\\
-i\sqrt{-y}\qquad &y\leq0
\end{array}
\right.
\end{eqnarray}

\begin{figure}[t]
\epsfysize=8cm
\epsffile{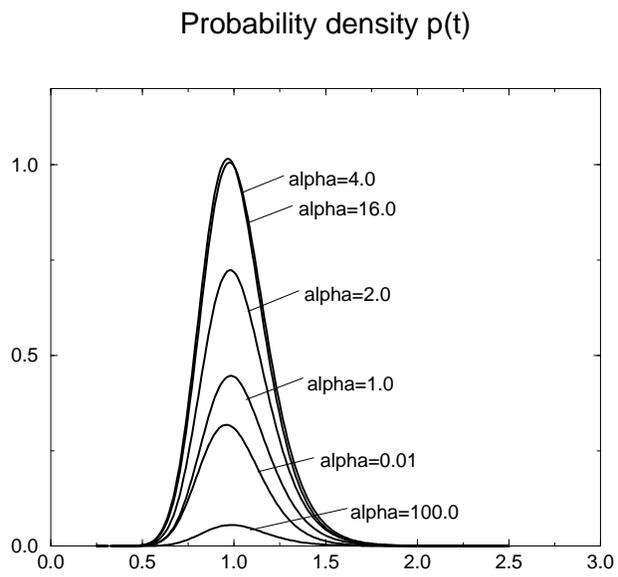}
\caption{Probability density of time of arrival for a Dirac's delta counter
placed at $x=0$, coupling constant alpha. The incoming wave packet starts at $t=0$,
$x=-4$, with velocity $v=4$}
\label{fig:fig2}
\end{figure}

\begin{figure}[b]
\epsfysize=8cm
\epsffile{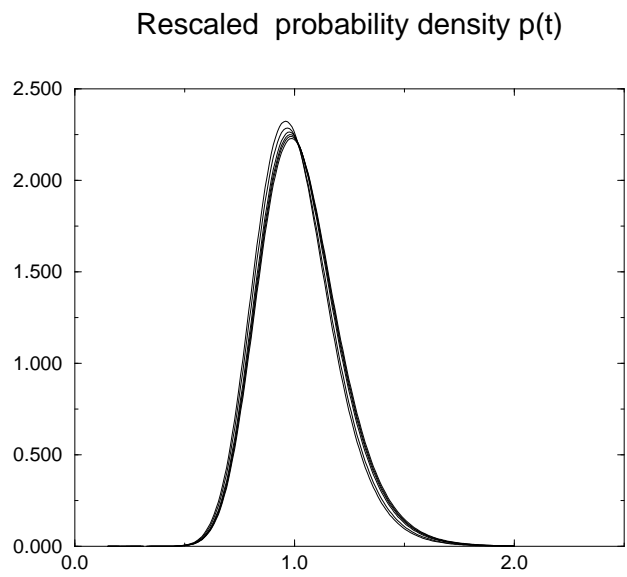}
\caption{Rescaled probability densities of Fig.1}
\label{fig:fig3}
\end{figure}

\begin{figure}[t]
\epsfysize=8cm
\epsffile{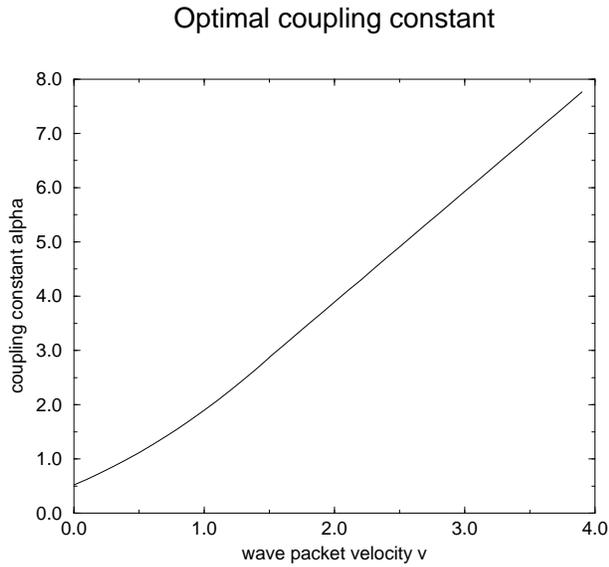}
\caption{Optimal coupling constant as a function of velocity of the incoming
wave packet. The dependence pretty soon saturates to a linear one. At the 
saturation value $P(\infty)\asymp 0.5.$}
\label{fig:fig4}
\end{figure}

\begin{figure}[b]
\epsfysize=8cm
\epsffile{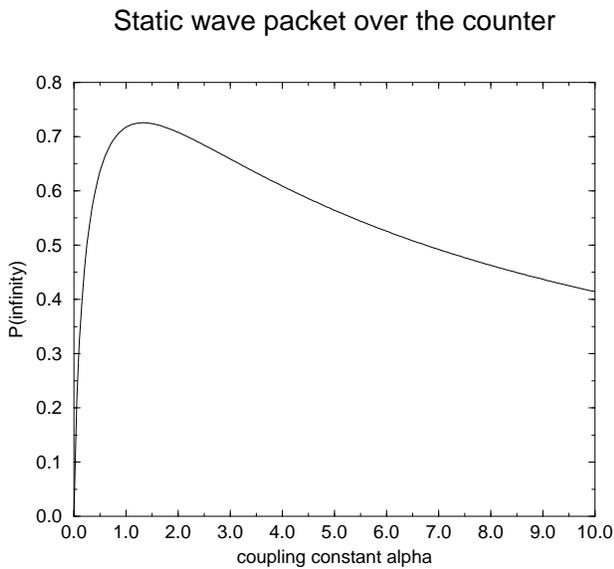}
\caption{$P(\infty)$ as a function of $\alpha$ for a static wave packet
centered over the counter. The maximum, of $P(\infty)=0.725448$ is reached
for $\alpha=1.3216 .$}
\label{fig:fig5}
\end{figure}

\begin{figure}[t]
\epsfxsize=3cm
\epsffile{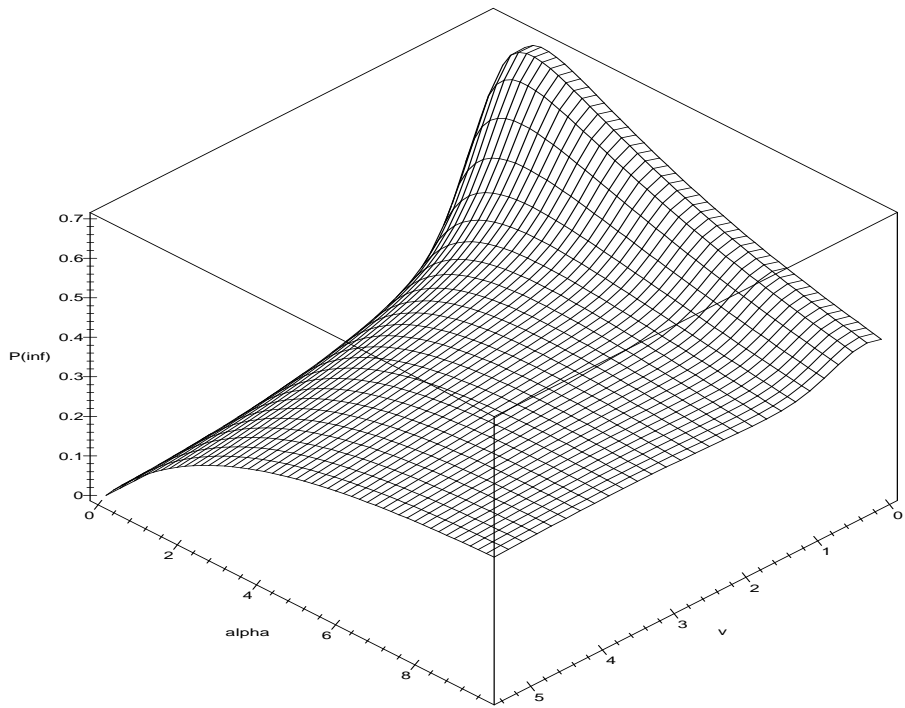}
\caption{$P(\infty )$ as a function of velocity $v$ and coupling constant $\alpha$ for
a static wave packet centered over the detector.}
\label{fig:fig6}
\end{figure}

The time of arrival probability curves of the counter for several values of
the coupling constant are shown in Fig.\ref{fig:fig2}. The incoming wave packet starts at 
$t=0$, $x=-4$, with velocity $v=4 .$ It is seen from the plot that the average
time at which the counter, placed at $x=0$, is triggered is about 
one time unit, independently of the value of the coupling constant. 
This numerical example shows that
our model of a counter serves can be used for measurements of time of
arrival. It is to be noticed that the shape of the response curve is 
almost insensitive to the value of the coupling constant. Fig.\ref{fig:fig3} shows
the curves of Fig.\ref{fig:fig2}, but rescaled in such a way that 
 the probability $P(\infty)=1$. The
only effect of the increase of the coupling constant in the interval 
$0.01-100$ is a slight shift of the response time to the left - which
is  intuitively clear. Notice that the shape of the curve in time corresponds
well to the shape of the initial wave packet in space.\\
For a given velocity of the packet
there is an optimal value of the coupling constant. In our dimensionless
units it is  $\alpha_{opt}\asymp 2v$.
Figure \ref{fig:fig4} shows this asymptotically linear dependence. At the optimal
coupling the total response probability $P(\infty)$ approaches the value $0.5 ,$
- the same as in the ultra--relativistic case.\\ 
By numerical calculations we have found that the maximal value of $P(\infty)$
that can be obtained for a single Dirac's delta counter and Schr\"odinger's
particle is slightly higher than $0.7 ,$ that corresponds to the value 
$\alpha=1.3$ of
the coupling constant. The dependence of $P(\infty)$ on the coupling constant
for a static wave packet (that is $v=0$) centered exactly over the detector is 
shown in Fig.\ref{fig:fig5}. 
Fig.\ref{fig:fig6} shows the dependence of $P(\infty)$ on both variables: $v$ and $\alpha .$ 

The value $0.7$ for the maximal response probability $P(\infty)$ of a detector
may appear to be rather strange. It is however connected with the point--like
structure of the detector in our simple model. For a composite detector, 
for instance already for a two--point detector, this restriction does not
apply and $P(\infty)$ arbitrarily close to $1.0$ can be obtained.
Our method applies as well to detectors continuously distributed in space.
In this case the efficiency of the detector (for a given initial wavepacket)
will depend on the shape of
the function $\Lambda(x)$. The absorptive complex potentials studied in 
\cite{muga94,muga95} are natural candidates for providing maximal efficiency as 
measured by $P(\infty)$ defined at the end of Sec. 2.2. 
\section{Concluding Remarks}
Our approach to the quantum mechanical measurement problem was
originally shaped to a large extent by the important paper by 
Klaus Hepp \cite{hepp72}.
He wrote there, in the concluding section: \lq The solution of the
problem of measurement is closely connected with the yet unknown
correct description of irreversibility in quantum mechanics...\rq
Our approach does not pretend to give an ultimate solution. But
it attempts to show that this "correct description" is, perhaps,
not too far away.\\
In the present paper we have only been able to scratch the surface of some 
of the new mathematical
techniques and physical ideas that are enhancing quantum theory in the framework
of EEQT, that free the quantum theory from the limitations of the standard
formulation. For a long time it was considered that quantum theory is only
about averages. Its numerical predictions were supposed to come {\em only}
from expectation values of linear operators. 
On the other hand in his 1973 paper \cite{wigner73} 
Wigner wrote: \lq It seems unlikely, 
therefore, that the superposition principle applies in full force to
beings with consciousness. If it does not, or if the linearity of
the equations of motion should be invalid for systems in which life
plays a significant role, the determinants of such systems
may play the role which proponents of the hidden variable theories
attribute to such variables. All proofs of the unreasonable nature
of hidden variable theories are based on the linearity of the 
equations ...\rq .
Weinberg \cite{weinberg89}
attempted to revive and to implement Wigner's idea of non-linear 
quantum mechanics.  He proposed a nonlinear framework and also
methods of testing for linearity. Warnings against potential dangers
of nonlinearity are well known, they were summarized in a recent paper 
by Gisin and Rigo \cite{gisin95}. The scheme of EEQT avoids these pitfalls
and presents a consistent and coherent theory. It
introduces necessary nonlinearity in the algorithm for generating
sample histories of individual systems, but preserves linearity
on the ensemble level. It is not only about averages but also
about individual events (cf. the event generating PDP algorithm of ref. 
\cite{blaja95a}). Thus it explains more, it predicts more and it opens
a new gateway leading beyond today's framework and towards new applications of 
Quantum Theory. These new applications may involve the problems
of consciousness. But in our opinion (supported in the all quoted papers
on EEQT, and also in the present one) quantum theory does not need 
neither consciousness nor human observers - at least not more
than any other probabilistic theory. On the other hand, to understand
mind and consciousness we may need Event Enhanced Quantum Theory. And more.\\
In the abstract to the present paper we stated that we "enhance elementary
quantum mechanics with three simple postulates". In fact the PDP
algorithm {\em replaces}\ the standard measurement postulates and enables 
us to {\em derive}\ them in a refined form. This is because EEQT {\em defines}\
precisely what {\em measurement}\ and {\em experiment}\ is - without
any involvement of consciousness or of human observers. It is only
for the purpose of the present paper - to introduce time observable
into elementary quantum mechanics as simply as possible - that we
have chosen to present our three postulates as postulates rather
than theorems. 
The time observable that we introduced
and investigated in the present paper is just one (but important)
trace of this nonlinearity.\footnote{This is why our time observable
does not fall into the family analysed axiomatically by Kijowski \cite{
kijowski74}.} 
Time of arrival, time of detector
response, is an "observable", is a random variable whose 
probability distribution function can be computed according
to the prescription that we gave in the previous section. But
its probability distribution is not a bilinear functional of
the state and as a result "time of arrival" can not be represented
by a linear operator, be it Hermitian or not. Nevertheless
our "time" of arrival is a "safe" nonlinear observable. Its
safety follows from the fact that what we called "postulates"
in the present paper are in fact "theorems" of the general
scheme EEQT. And EEQT is the {\em minimal} extension of quantum
mechanics that accounts for events: no extra unnecessary
hidden variables, and linear Liouville equation for ensembles.\\
Our definition of time of arrival bears some similarity to
the one proposed long ago by Allcock \cite{allcock69}. 
Although we disagree in several important points with
the premises and conclusions of this paper, nevertheless
the detailed analysis of some aspects of the problem
given by Allcock was prompting us to formulate and to solve 
it using the new perspective and the new tools that EEQT 
endowed us with. Our approach to the problem of time of arrival 
goes in a similar direction as 
the one discussed in an (already quoted) interesting recent 
paper by Muga and co--workers \cite{muga95}. We share many of his views. The 
difference being  that what the authors of \cite{muga95} call
"operational model" we promote to the role of a fundamental
new postulate of quantum theory. We justify it and point out that
it is a theorem of a more fundamental theory - EEQT.
Moreover we take the non--unitary evolution before the detection 
event seriously and point out that the new theory is experimentally 
falsifiable.  \\   
Once the time of arrival observable has been defined, it is
rather straightforward to apply it. In particular our time
observable solves Mielnik's "waiting screen problem" 
\cite{mielnik94}. But not only that; with our
{\bf precise} definition at hand,  one can approach again the old 
puzzle of time--energy uncertainty relation in the spirit 
of Wigner's analysis \cite{wigner72} (cf. also \cite{recami77,pfeifer94}. 
One can also approach
afresh the other old problem: that of decay times (see \cite{
horwitz93} and references therein) and of  
tunneling times (\cite{recami92,landauer94,leavens95} and 
references therein). 
This last problem needs however more than just one detector. 
We need to analyse the joint distribution probability for two 
separated detector. We must also know how to describe
the unavoidable disturbance of the wave function when
the first detector is being triggered. For this the
simple postulates of this paper do not suffice. But
the answer is in fact quite easy if using the event
generating algorithm of EEQT.\\
More investigations needs also our "shadowing 
effect" of section 2.3. Every "real" detector acts
not only as an information exchange channel, but also
as an energy--momentum exchange channel. Every real
detector has not only its "information temperature"
described by our coupling constant $\lambda$ (cf. Sec.
2.1), but also ordinary temperature. Experiments
to test the effect must take care in separating
these different contributions to the overall
phenomenon. This is not easy. But the theory is falsifiable
in the laboratory and critical experiments might
be feasible within the next couple of years.\\
In the introductory chapter the problem of 
extension of the present framework to the
relativistic case has been shortly mentioned. Work
in this direction is well advanced and we
hope to be able to report its result soon. But
this will not be end of the story. At the very 
least we have much to learn about the nature and the 
mechanism of the coupling between $Q$ and $C$.\footnote
{More comments in this direction can be found in 
\cite{blaja95a} and also under WWW address
of the Quantum Future Project: http://www.ift.uni.wroc.
pl/\ $\tilde{}$ajad/qf.htm}
 
\vskip10pt
\noindent
{\bf Acknowledgements}\\
One of us (A.J) acknowledges with thanks support of A. von Humboldt
Foundation. He also thanks Larry Horwitz for encouragement, 
Rick Leavens for his interest and pointing out the relevance of
Muga's group papers and to Gonzalo Muga for critical comments.
We are indebted to Walter Schneider for his 
interest, critical reading of the manuscript and for supplying 
us with relevant informations. We thank Rudolph Haag for sending 
us the first draft of \cite{haag96}.

\newpage
\end{document}